%% file: main.tex
\ifdefined\pdfsuppressptexinfo\pdfsuppressptexinfo=15\relax\fi
\ifdefined\pdfinfoomitdate\pdfinfoomitdate=1\relax\fi
\ifdefined\pdftrailerid\pdftrailerid{}\fi
\documentclass[11pt,a4paper,logo,copyright,nonumbering]{xiaomi}
\usepackage[authoryear,round]{natbib}
\usepackage{latexsym}

\input{math_commands.tex}

\usepackage{graphicx}
\usepackage{capt-of}
\usepackage{booktabs}
\usepackage{amssymb}
\usepackage{xcolor,colortbl}
\usepackage{tabularx}
\usepackage{wrapfig}
\usepackage{stfloats}
\usepackage{array,tikz,tcolorbox,placeins}
\tcbuselibrary{breakable}
\usetikzlibrary{arrows.meta,positioning}
\definecolor{ink}{HTML}{34313B}
\definecolor{purple}{HTML}{7645A6}
\definecolor{pale}{HTML}{F4F1F7}
\definecolor{blue}{HTML}{527F9A}
\definecolor{warm}{HTML}{AD6749}
\definecolor{rulegray}{HTML}{D8D5DC}
\newcolumntype{Y}{>{\raggedright\arraybackslash}X}
\newenvironment{casebox}[1]{\begin{tcolorbox}[breakable,colback=pale,colframe=rulegray,boxrule=.4pt,arc=1pt,left=7pt,right=7pt,top=5pt,bottom=5pt,title={#1},colbacktitle=pale,coltitle=ink,fonttitle=\bfseries\small]\setlength{\parskip}{4pt}\raggedright}{\end{tcolorbox}}

\usepackage{hyperref}
\hypersetup{pdfcreator={},pdfproducer={}}
\usepackage{url}

\definecolor{tablegroup}{HTML}{EFF7EF}
\definecolor{tableours}{HTML}{F6F0E4}
\definecolor{tablemetric}{HTML}{F2F2F2}

\DeclareRobustCommand{\method}{\mbox{\textsc{CodeSpecWeave}}}
\DeclareRobustCommand{\bench}{\mbox{\textsc{CodeSpecWeave}}-\allowbreak\mbox{\textsc{Bench}}}
\graphicspath{{figures/}}

\title{When Requirements Interact:\\
Scaling Long-Horizon Coding via Interdependent Specifications}
\author{Yishuo Cai, Minzheng Wang, Liang Zhao, Linghao Zhang, Lei Li, Shicheng Li, Xu Sun, Fuli Luo}
\makeatletter
\renewcommand{\@author}{%
  \centering\normalfont\fontsize{11}{15}\selectfont
  Yishuo Cai\textsuperscript{1,2,*,\textdaggerdbl} \quad Minzheng Wang\textsuperscript{1,*} \quad Liang Zhao\textsuperscript{1} \quad Linghao Zhang\textsuperscript{1} \\[4pt]
  Lei Li\textsuperscript{1,3} \quad Shicheng Li\textsuperscript{1} \quad Xu Sun\textsuperscript{2,\textdagger} \quad Fuli Luo\textsuperscript{1,\textdagger} \\[10pt]
  \textsuperscript{1}LLM Core, Xiaomi \quad \textsuperscript{2}Peking University \\[4pt]
  \textsuperscript{3}University of Hong Kong\par
}
\makeatother

\begin{abstract}
\input{sections/abstract}
\end{abstract}

\begin{document}
\maketitle
\begingroup
\renewcommand{\thefootnote}{}
\footnotetext[0]{\textsuperscript{*}Equal contribution. \textsuperscript{\textdagger}Corresponding authors. \textsuperscript{\textdaggerdbl}Work done during an internship at Xiaomi.}
\endgroup

\input{sections/introduction}

\input{sections/preliminaries}
\input{sections/method}

\input{sections/experiments}

\FloatBarrier
\input{sections/related_work}
\input{sections/conclusion}

\bibliography{references}

\clearpage
\appendix
\input{sections/appendix}

\end{document}

%% file: math_commands.tex
\usepackage{amsmath,amsfonts,bm}

\def\eqref#1{equation~\ref{#1}}

\def\1{\bm{1}}

\DeclareMathAlphabet{\mathsfit}{\encodingdefault}{\sfdefault}{m}{sl}
\SetMathAlphabet{\mathsfit}{bold}{\encodingdefault}{\sfdefault}{bx}{n}

%% file: sections/abstract.tex
To succeed in real-world long-horizon software development, coding agents must satisfy multi-dimensional task specifications, such as interface compatibility and error handling, and do so without breaking existing dependencies. However, current works lack specification in both quantity and diversity, often emphasizing primary goals while ignoring complex constraints. Synthesizing tasks to bridge this gap presents two challenges: naively adding specifications risks reducing tasks to trivial checklists rather than compelling joint reasoning, and verifying these subtle interactions risks overlooking hidden flaws or rejecting valid alternatives. To address these challenges, we introduce \method{}, a framework for synthesizing tasks with rich, interdependent specifications. Grounded in real developer seeds, a synthesis agent iteratively develops both coupled task requirements and executable tests, while a review agent audits solution trajectories from multiple models to guarantee test reliability. Using this pipeline, we construct 4,670 training tasks across 2,865 repositories. Applying Reinforcement learning on these tasks significantly improves performance over the base model on DeepSWE-v1.1, largely outperforming existing baselines. Furthermore, we curate \bench{}, a rigorous benchmark comprising 100 tasks and 1,044 specifications, which clearly separates frontier models and reveals significant headroom in handling interdependent engineering constraints.

%% file: sections/introduction.tex
\section{Introduction}
\label{sec:introduction}

Recent advances in large language models (LLMs) have enabled coding agents to explore repositories, modify code, execute tests, and iteratively refine their solutions~\citep{sweagent,swebench,swebench_pro,tan2026swe}. As these agents navigate the increasing complexity of real-world software architectures, they must implement changes without breaking behavior that other components already depend on~\citep{swebench,swebench_pro}. In practice, these expectations span multiple requirements, such as interface compatibility, state consistency, and error handling~\citep{archcode,specrover}. Together, they form concrete task specifications that an agent must satisfy throughout exploration, coding, and testing.

\begin{figure*}[t]
\centering
\includegraphics[width=5.5in]{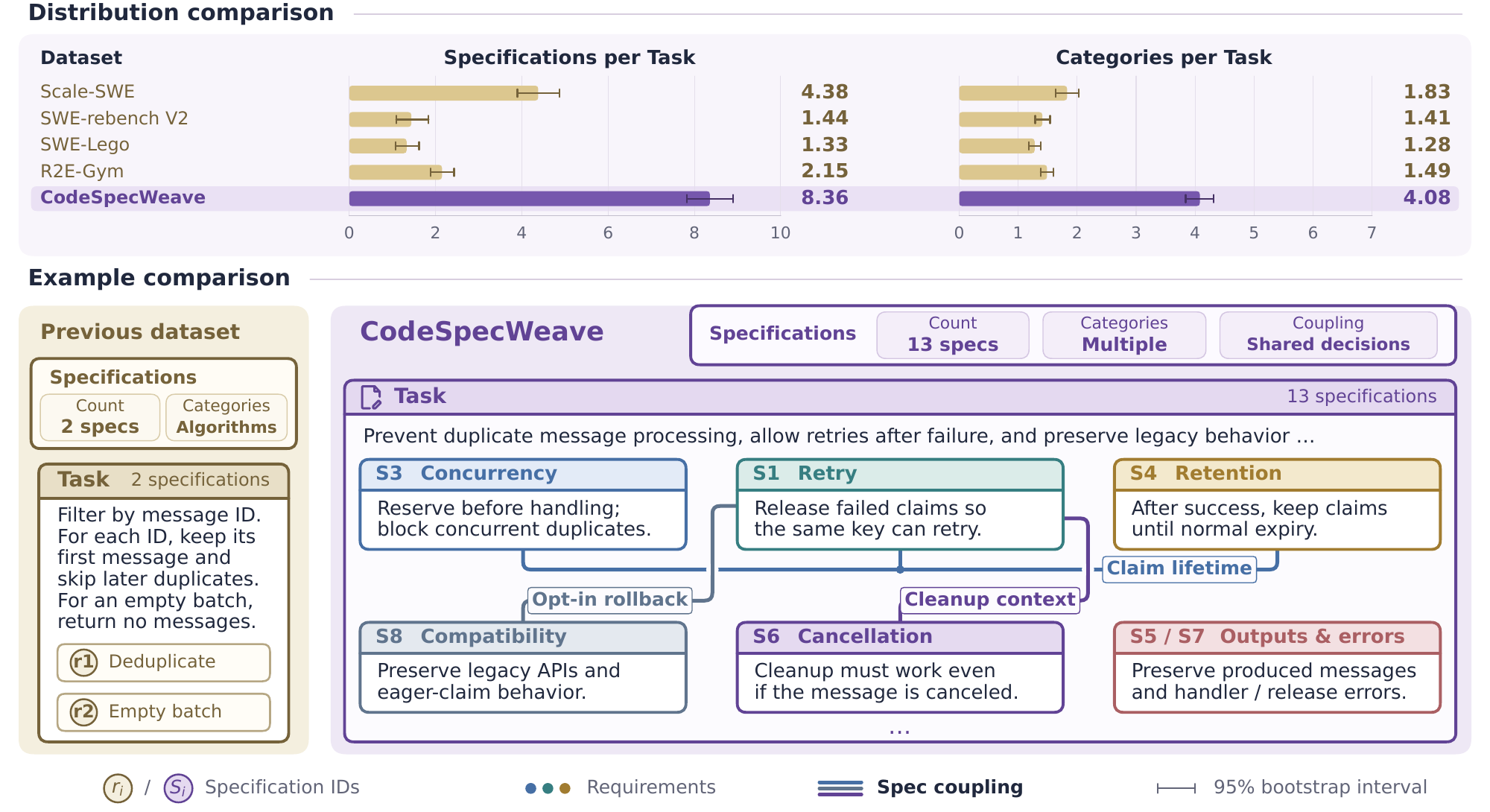}
\caption{\textbf{Comparison of task specifications across datasets.}
\textbf{Top:} Mean extracted specifications and distinct behavioral categories per task across existing SWE training datasets and \method{}, with 95\% bootstrap confidence intervals. Category counts follow the taxonomy in Table~\ref{tab:spec-categories}.
\textbf{Bottom:} Example task structures. Prior datasets typically contain few independent rules (left), whereas \method{} tasks feature multiple interdependent specifications tied together by shared execution contexts and lifecycle constraints (right).}
\label{fig:overview}
\end{figure*}

Developing this capability requires a large number of executable tasks that encompass these diverse engineering requirements. Existing works typically construct executable tasks by mining issues and pull requests~\citep{swegym_icml}, with limited attention to the role and properties of task specifications. Our empirical analysis reveals that existing data fall significantly short in both specification density and structural diversity (Figure~\ref{fig:overview}, top), often emphasizing the primary development goal while leaving compatibility, state consistency, and error handling unaddressed. Consequently, agents trained on such partial specifications often fail in practice, as they struggle to satisfy the intricate constraints inherent in real-world repositories. This deficiency underscores the need for a scalable pipeline to synthesize tasks grounded in realistic and diverse specifications.

However, synthesizing such tasks presents two connected challenges. First, purely adding specifications may inadvertently reveal solution paths~\citep{taskpilot} or reduce the task to a checklist of independent edits. Specification design should therefore organize requirements into interdependent specifications where individual implementation choices constrain one another, compelling agents to coordinate across requirements through joint reasoning to achieve globally coherent implementations (Figure~\ref{fig:overview}, bottom). Second, verifying such interdependent specifications is uniquely challenging: the subtle interactions among requirements make it easy for tests either to miss hidden flaws~\citep{swemutation,coharden} or to reject valid solutions~\citep{swebench_verified}. Reliable verification requires strict alignment with both individual requirements and their interactions, while accommodating different implementations that satisfy them.

To this end, we introduce \method{}, a framework for synthesizing coding tasks with rich, interdependent specifications. Grounded in specification seeds derived from real developers, a synthesis agent explores the repository context and develops the task requirements and executable test suites together in an iterative loop, intentionally designing coupled constraints that require simultaneous satisfaction while refining the tests to strictly align with them. To ensure test reliability, a review agent further examines the solution trajectories and patches from multiple models, filtering out tests that accept incomplete implementations or reject valid alternatives. Using this pipeline, we construct 4,670 training tasks across 2,865 repositories. Applying Reinforcement learning on these tasks significantly improves performance over the base model on DeepSWE-v1.1. In addition, we curate \bench{} through careful manual review, compiling 100 tasks with 1,044 specifications to evaluate coding agents under interdependent specifications. Benchmarking leading models clearly separates their capabilities, revealing shared failure modes and significant room for improvement.

Our contributions are as follows:\begin{itemize} 
    \item We introduce \method{}, a synthesis framework that builds on real developer specification seeds to generate coding tasks with interdependent specifications and reliable verifiers.
    \item We construct a scalable training set of 4,670 tasks across 2,865 repositories and demonstrate its effectiveness through reinforcement learning, yielding substantial performance gains on DeepSWE-v1.1 over strong training baselines.
    \item We present \bench{}, a benchmark comprising 100 tasks with 1,044 specifications. Evaluations across leading coding agents reveal notable performance gaps and substantial headroom under interdependent requirements.
\end{itemize}

%% file: sections/method.tex
\section{\method}
\label{sec:method}

\begin{figure*}[t]
\centering
\includegraphics[width=5.5in]{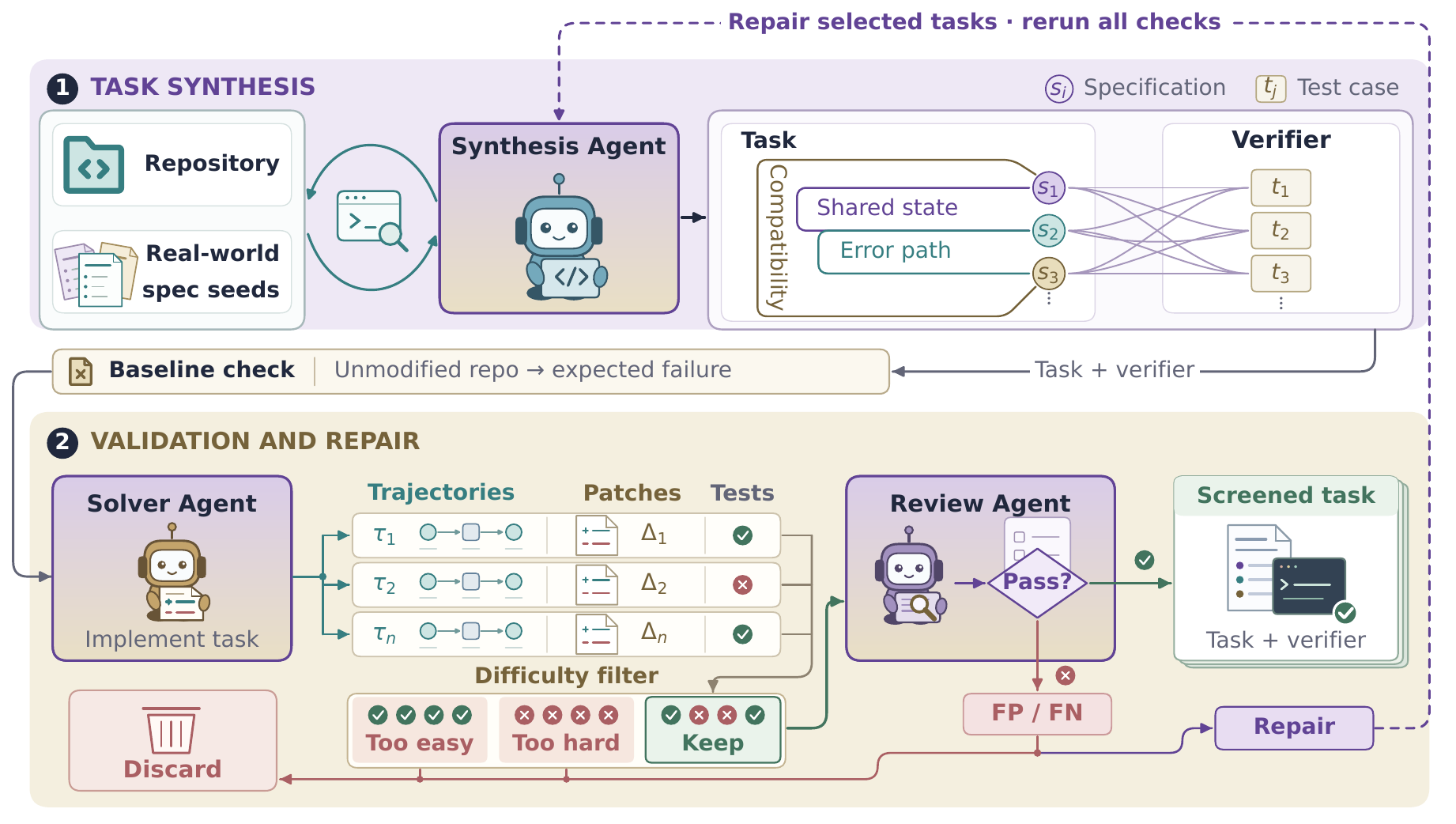}
\caption{\textbf{Overview of \method.} 
The synthesis agent repeatedly consults specification seeds, explores the repository, and runs generated tests to refine a task and its verifier. Schematic links between specifications and test cases illustrate many-to-many coverage. Solver trials and trajectory review then assess difficulty and query--test alignment. Accepted task--verifier pairs are retained; selected candidates with false positives or false negatives are repaired and revalidated.}
\label{fig:pipeline}
\end{figure*}

\method\ transforms requirements observed in real software development into coding tasks, addressing two connected challenges: compelling agents toward holistic exploration and joint reasoning under rich specifications, and scaling reliable test verification aligned with stated requirements.
We first introduce the task formulation (Section~\ref{sec:preliminaries}) and design principle (Section~\ref{sec:design-principle}).
The framework consists of two core stages (Figure~\ref{fig:pipeline}):
\emph{task synthesis} constructs interdependent requirements and linked tests guided by specification seeds (Section~\ref{sec:task-synthesis});
\emph{validation and repair} reviews trajectories generated by multiple model trials to identify and repair misalignments between queries and tests (Section~\ref{sec:validation-repair}).

\subsection{Preliminaries and Problem Formulation}
\label{sec:preliminaries}

We represent a task as a tuple $\mathcal{T}=(\mathcal{R},\mathcal{Q},\mathcal{S},\mathcal{E})$.
Here, $\mathcal{R}$ is the code repository, $\mathcal{Q}$ is the natural-language request provided to the solver, $\mathcal{S}=\{s_1,\ldots,s_m\}$ is the set of specifications embedded within $\mathcal{Q}$ that a valid patch must satisfy, and $\mathcal{E}$ is an execution environment equipped with a verifier $V$.
Given $\mathcal{Q}$ and $\mathcal{R}$, a solver produces a patch $\Delta$ through repository interactions.
We write $s_i(\Delta)=1$ when the patched repository satisfies specification $s_i$, and $0$ otherwise.
The set of valid patches is defined as:
\begin{equation}
\mathcal{F}(\mathcal{T})=\{\Delta:\ s_i(\Delta)=1\ \text{for all }i\}.
\end{equation}
In practice, the verifier $V$ checks whether $\Delta \in \mathcal{F}(\mathcal{T})$ by executing test suites.

\subsection{Design Principle: Interdependent Specifications}
\label{sec:design-principle}

Integrating rich specifications into coding tasks requires organizing how requirements relate to one another, rather than merely accumulating them.
In our initial synthesis attempts, we observed that when relying solely on specification count, the synthesized tasks remained trivial: language models achieved high success simply by satisfying each requirement in isolation.
This observation motivates distinguishing \emph{additive specifications}, whose requirements admit largely separate decisions, from \emph{interdependent specifications}, where a choice made to satisfy one requirement constrains the feasible options for another.

Formally, let design variables $d = (d_1, \ldots, d_k)$ denote the engineering choices permitted by specification set $\mathcal{S}$ within feasible space $\mathcal{F}_d$.
An idealized additive task factorizes independently as $\mathcal{F}_d = \prod_j \mathcal{F}_j$, reducing to a step-by-step checklist where each choice can be resolved greedily in isolation.
In contrast, interdependent specifications break this factorization ($\mathcal{F}_d \subsetneq \prod_j \mathcal{F}_j$), such that selecting one choice strictly restricts the feasible alternatives for others.
Because local edits cannot be decoupled, the solver must reason across all requirements simultaneously, presenting much more challenging tasks for existing agents.

\begin{figure*}[t]
\centering
\includegraphics[width=5.5in]{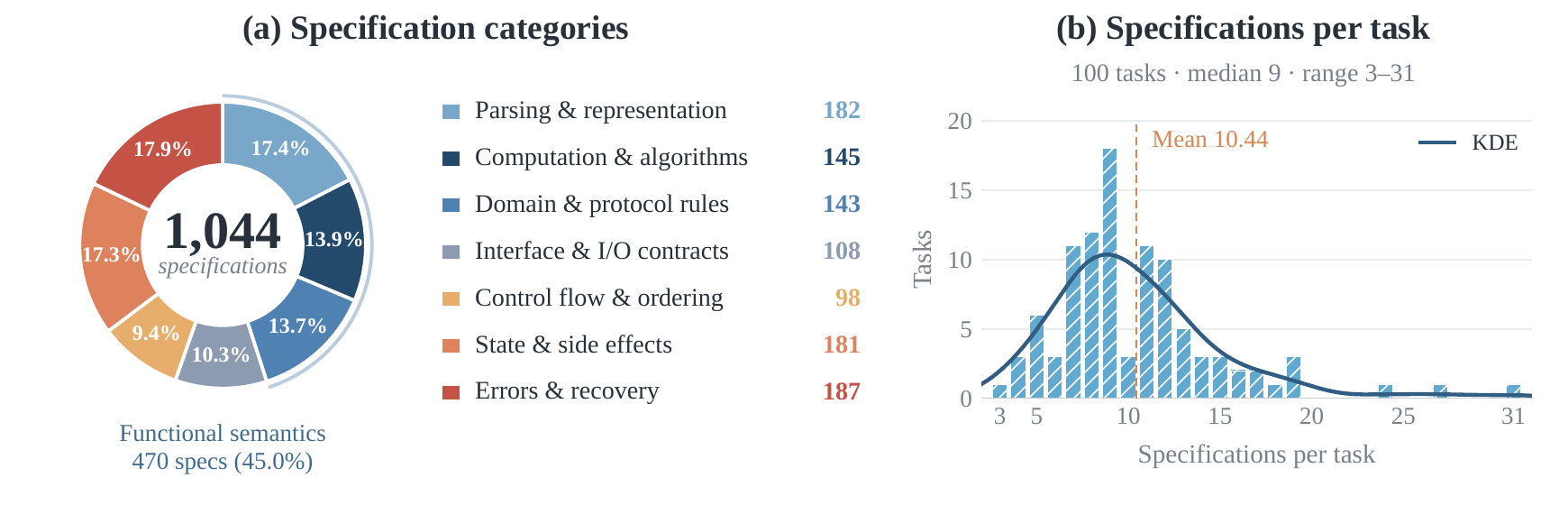}
\caption{\textbf{Specification distribution in \bench.} (a) The five behavioral categories in Table~\ref{tab:spec-categories}, with functional semantics expanded into three blue subcategories. Each of the 1,044 specifications contributes once to one of the seven segments. (b) Specifications per task across 100 tasks (mean: 10.44).}
\label{fig:benchmark-composition}
\label{fig:benchmark-spec-groups}
\label{fig:benchmark-profile}
\end{figure*}

\subsection{Task Synthesis}
\label{sec:task-synthesis}

To reflect real-world developer requirements, we extract specification seeds from open-source repositories by mining commit histories, issue discussions, and pull requests.
After filtering out conversational chatter and raw code fragments, normalized deduplication yields 104,182 candidate specification units across six natural languages.
These seeds inform a compact specification reference containing examples, definitions, and boundary guidance for task synthesis.

Given a repository $\mathcal{R}$, specification seeds, and prompt instructions to construct interdependent specifications, the synthesis agent explores the codebase to formulate development tasks.
It combines complementary requirements, anchoring them to the same state transitions or core components so that satisfying one requirement constrains the implementation of others.
For instance, a task may couple input validation, transactional rollback, and backward compatibility for existing readers, ensuring that satisfying any single requirement in isolation still violates the broader contract.
The agent iteratively drafts the task request $\mathcal{Q}$, fine-grained specifications $\mathcal{S}$, and corresponding test cases $\mathcal{A}$.
These tests map flexibly to specifications, allowing multiple checks to validate a single requirement or an integration test to cover several obligations simultaneously.
Throughout this process, the agent executes generated tests directly in the repository and refines them based on runtime feedback.

\subsection{Validation and Repair}
\label{sec:validation-repair}

\input{tables/query_test_alignment}

Simultaneously generating interdependent specifications and tests does not guarantee capturing the intended requirement boundary.
Because multiple requirements closely interact, tests often fail to check flaws that arise only when the requirements interact at runtime, allowing incorrect solutions to pass~\citep{swemutation,coharden}.
At the same time, verifying coupled requirements easily causes tests to assume a specific implementation, rejecting valid solutions that satisfy the full specification~\citep{swebench_verified}.
Our pilot generation reflects this difficulty: automated reviews identified alignment flaws in 2,267 of 3,067 candidate tasks (73.9\%; Table~\ref{tab:query-test-alignment}), confirming that complex specifications make reliable test generation exceedingly hard.

Because these flaws only appear when code actually runs against the requirements, inspecting static code or tests alone is insufficient.
Instead, we collect candidate patches $\Delta$ and their execution trajectories $\tau$ from independent coding models across multiple attempts, filtering for tasks with mixed solver success to keep challenging yet solvable tasks.
Comparing these execution trajectories with test results reveals errors that are invisible during generation: runs that violate multiple requirements expose missing checks, while valid runs that fail indicate tests that enforce unintended restrictions.
A review agent inspects the full tuple $(\mathcal{Q}, \mathcal{A}, \Delta, \tau)$ to iteratively reconcile the tests with the interdependent specifications.
In one generation batch of 14,905 initial tasks, difficulty filtering retained 3,275 candidates, from which this review approved 766 verified tasks.

For selected candidates with identified false positives or false negatives, a repair agent uses the review findings and execution trajectories to revise the task request or verifier.
Each repaired task--verifier pair then undergoes execution checks, fresh solver trials, and alignment review before acceptance.

\input{sections/benchmark_construction}

%% file: tables/query_test_alignment.tex
\begin{wraptable}{R}{0.44\textwidth}
\vspace{-8pt}
\centering
\small
\begin{tabular}{@{}lrr@{}}
\toprule
Issue & Tasks & \% \\
\midrule
False positives only & 1,548 & 50.5 \\
False negatives only & 426 & 13.9 \\
Both & 293 & 9.6 \\
None observed & 800 & 26.1 \\
\midrule
Total with issues & 2,267 & 73.9 \\
\bottomrule
\end{tabular}
\caption{Test alignment breakdown on 3,067 candidates.}
\label{tab:query-test-alignment}
\end{wraptable}

%% file: sections/benchmark_construction.tex
Using the full \method{} pipeline, we synthesize 4,670 training tasks across 2,865 repositories for reinforcement learning.
For evaluation, we curate \bench{}, a benchmark of 100 manually reviewed tasks.
We provide details of the manual review in Appendix~\ref{app:author-review}.
Figure~\ref{fig:benchmark-composition} summarizes its 1,044 specifications by category and task, with a mean of 10.44 specifications per task. The models and harness used for data construction are listed in Appendix~\ref{app:agents}.

%% file: sections/experiments.tex
\section{Experiments}
\label{sec:experiments}

\input{sections/benchmark}

\subsection{Training on \method}
\label{sec:rl-utility}

\subsubsection{Training and evaluation setup}

\paragraph{Models.}
We apply reinforcement learning to two Mixture-of-Experts (MoE) models, MiMo-V2.5~\citep{mimo_v25} and MiMo-V2.6-Flash-SFT~\citep{mimo_v26}, each with 310B total and 15B active parameters.

\paragraph{Training datasets.}
We compare three RL datasets: \method{}, SWE-PI, and MiMo Coding.
\method{} contains 4,670 tasks across 2,865 repositories. SWE-PI is a mixture of SWE datasets released by Prime Intellect~\citep{primeintellect2026scalingagenticrl}, comprising 30,128 tasks from Scale-SWE~\citep{scaleswe}, SWE-Lego~\citep{swelego}, SWE-rebench V2~\citep{swerebench}, and Multi-SWE-RL~\citep{multiswebench}.
MiMo Coding~\citep{mimo_v26} is curated from diverse collection pipelines, including issue-driven development, everyday development workflows, functionality implementation, and long-horizon engineering.
Training hyperparameters are provided in Appendix~\ref{app:training}.

\input{tables/rl_comparison}
\begin{figure*}[t]
\centering
\includegraphics[width=6.0in]{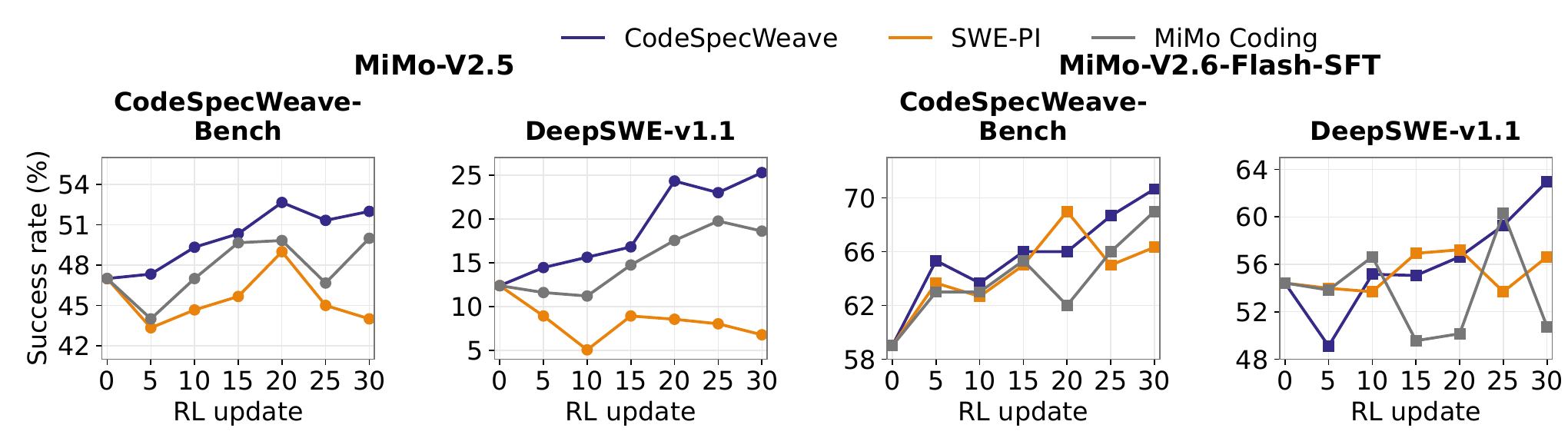}
\caption{Learning curves for \method{}, SWE-PI, and MiMo Coding, trained from MiMo-V2.5 (left) and MiMo-V2.6-Flash-SFT (right).}
\label{fig:learning}
\end{figure*}

\paragraph{Evaluation benchmarks.}
We evaluate on four coding benchmarks.
\bench{} contains 100 tasks from 100 repositories with interdependent specifications.
DeepSWE-v1.1 contains 113 long-horizon coding tasks~\citep{datacurve_v11}.
SWE-bench Verified comprises 500 human-validated GitHub issues from Python repositories~\citep{swebench,swebench_verified}.
SWE-bench Pro provides 731 public tasks across multiple programming languages~\citep{swebench_pro}.
Evaluation settings are provided in Appendix~\ref{app:rl-matrix}.

\subsubsection{Evaluation results}

Table~\ref{tab:rl-matrix} compares performance across four benchmarks.
Across both MiMo-V2.5 and MiMo-V2.6-Flash-SFT backbones, \method{} yields the highest average scores (50.15\% and 67.60\%), outperforming other training datasets. The improvements are especially clear on \bench{} and DeepSWE. Figure~\ref{fig:learning} shows the RL updates of \method{} across both backbones on these two benchmarks. On \bench{}, \method{} consistently achieves the top scores across both base models (52.00\% and 70.67\%). The advantage is even more pronounced on DeepSWE, which presents challenging long-horizon coding tasks. While other datasets show limited gains or even negative transfer (e.g., SWE-PI on MiMo-V2.5), \method{} substantially boosts scores to 25.30\% and 62.98\%. Meanwhile, \method{} maintains competitive performance on SWE-bench Verified and SWE-bench Pro.

\input{tables/training_behavior}
\paragraph{Model behavior.}
After \method{} training, MiMo-V2.5 generates 4.8 times as many thinking tokens and MiMo-V2.6-Flash-SFT 3.0 times as many (Table~\ref{tab:training-behavior}).
Both models also modify significantly more files and lines on average, indicating a shift from localized patches toward cross-file refactoring to accommodate interdependent specifications.
Crucially, this expanded modification scope is grounded in active verification: training increases the number of edit--test loops per trajectory (from 3.48 to 4.34 for MiMo-V2.5, and from 2.96 to 4.90 for Flash-SFT).
Each loop consists of one or more code edits followed by executable test feedback in a subsequent turn.
With an increased proportion of trajectories sustaining repeated loops ($\ge 2$ loops reaching over 70\%), this demonstrates that \method{} trains agents to systematically navigate and validate complex constraints without breaking existing dependencies through tight, iterative feedback loops.

\subsubsection{Performance by specification count and category}
\label{sec:specification-performance}

\begin{figure}[!t]
\centering
\includegraphics[width=\columnwidth]{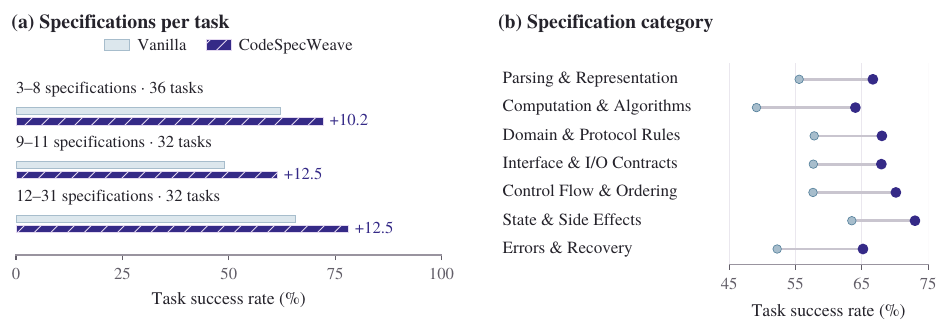}
\caption{Task success rates on \bench{} for MiMo-V2.6-Flash-SFT before and after reinforcement learning on the 4,670-task \method{} training set. Results cover all 100 benchmark tasks. (a) Tasks grouped by specifications per task, with annotations showing percentage-point gains over Vanilla. (b) Tasks grouped by specification category (each point denotes the success rate on tasks requiring that category).}
\label{fig:specification-profiles}
\end{figure}

\method{} improves Flash-SFT across tasks with different numbers of specifications, with gains of 10.2--12.5 percentage points (Figure~\ref{fig:specification-profiles}).
The gain is greatest in both the 9--11 and 12--31 specification groups, suggesting that \method{} helps the model better handle tasks with many interacting requirements.
Gains are also broadly comparable across the seven categories in Table~\ref{tab:spec-categories} (9.5--14.9 points), covering both functional requirements and constraints on interfaces, execution order, state, and error handling.

\subsubsection{Training dynamics}

\begin{figure*}[t]
\centering
\includegraphics[width=5.5in]{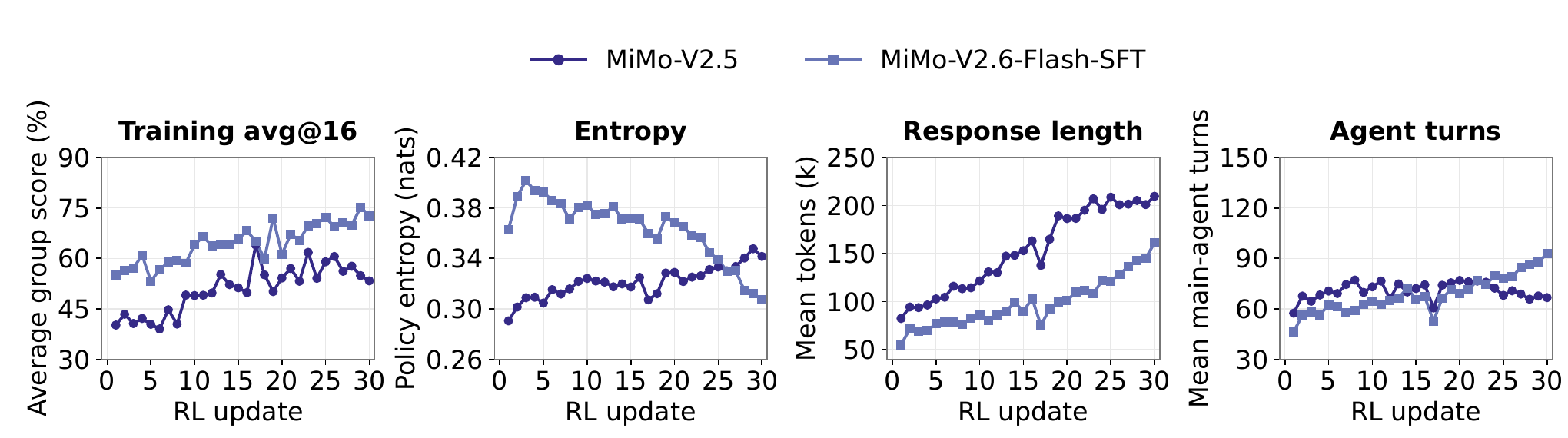}
\caption{Training dynamics through RL update for the two \method{} runs. From left to right: mean task-group score, policy entropy, response length, and main-agent turns.}
\label{fig:training-dynamics}
\end{figure*}

Figure~\ref{fig:training-dynamics} illustrates the training dynamics across both backbones.
Task scores rise steadily across updates while policy entropy remains stable (0.29--0.40), confirming consistent policy improvement.
To resolve interdependent specifications, the two models exhibit distinct adaptation strategies: MiMo-V2.5 scales its single-turn effort through extensive elaboration (more than doubling its response length), whereas Flash-SFT relies on active multi-turn interaction by nearly doubling its agent turns.
These trajectories demonstrate that \method{} reinforces sustained problem solving and environment exploration.

%% file: sections/benchmark.tex
\subsection{Evaluation on \bench}
\label{sec:benchmark}

\begin{figure*}[t]
\centering
\includegraphics[width=5.5in]{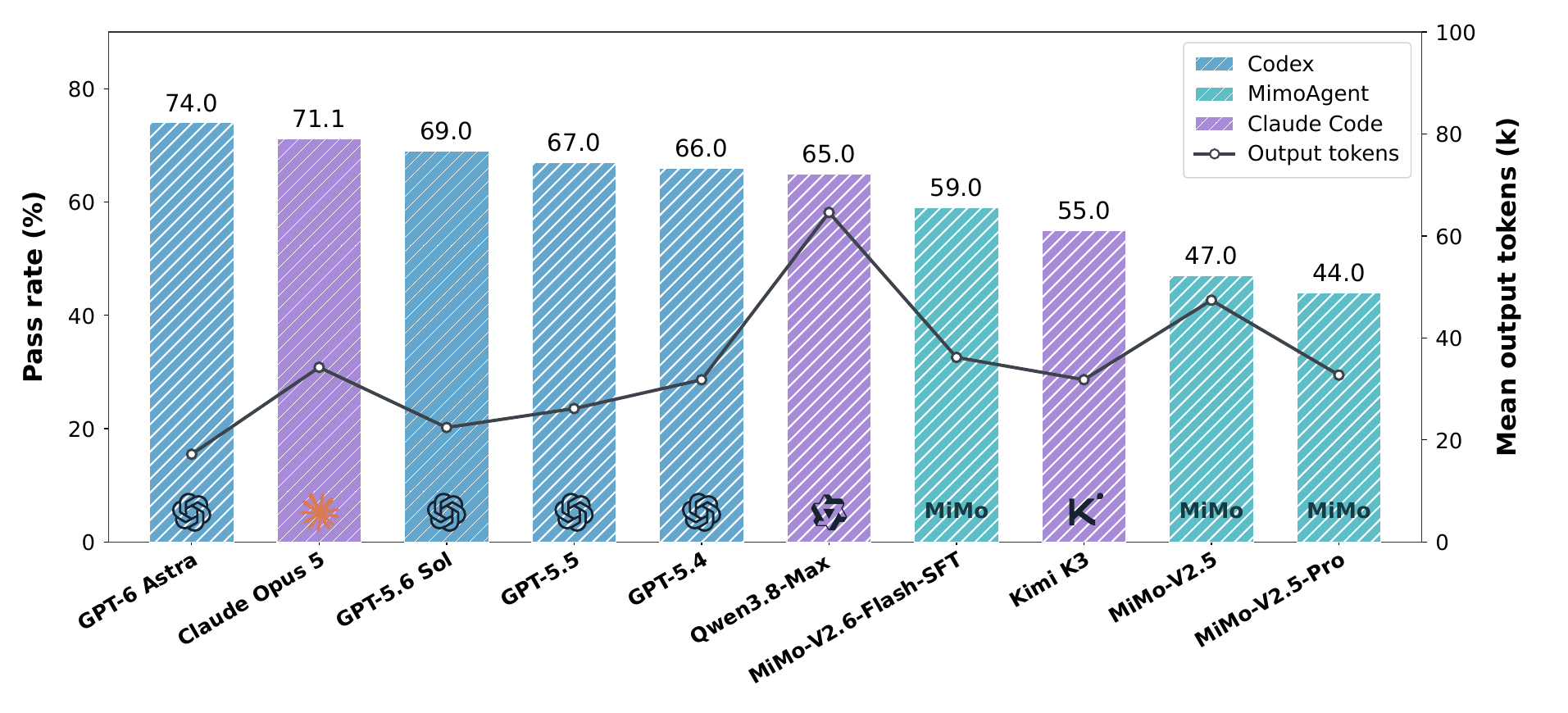}
\caption{\textbf{Pass rates and mean reported output tokens on \bench.}
Hatched bars show pass rates (left axis); the line shows mean output tokens (right axis). Colors identify agent frameworks.}
\label{fig:benchmark-results}
\end{figure*}

\paragraph{Evaluation setup.}
We evaluate GPT-5.4, GPT-5.5, GPT-5.6 Sol, and GPT-6 Astra~\citep{gpt54,gpt55,gpt56sol,gpt6astra} with Codex~\citep{codex_cli};
Claude Opus 5, Kimi K3, and Qwen3.8-Max~\citep{claude_opus5,kimi_k3,qwen38max} with Claude Code~\citep{claude_code};
and MiMo-V2.5, MiMo-V2.5-Pro~\citep{mimo_v25,mimo_v25_pro}, and MiMo-V2.6-Flash-SFT with MimoAgent~\citep{mimo_v26}.

\paragraph{Results.}
Figure~\ref{fig:benchmark-results} shows performance across representative agents on \bench{}.
Pass rates range from 44.0\% to 74.0\%, demonstrating that the benchmark clearly distinguishes model capabilities while leaving significant headroom, as even the strongest model (GPT-6 Astra at 74.0\%) fails on 26\% of the attempts.

Higher task success does not require more output tokens.
GPT-6 Astra achieves the highest pass rate with the lowest mean output length (17.2k tokens).
Within Codex, pass rate decreases monotonically with mean output length across the four GPT models, from 74.0\% at 17.2k tokens to 66.0\% at 31.8k.
Within Claude Code, Claude Opus~5 exceeds Qwen3.8-Max by 6.1 percentage points while producing approximately 47\% fewer output tokens, indicating that extra length largely reflects unproductive trial-and-error when resolving interdependent specifications.

\input{sections/benchmark_failure_analysis}

%% file: sections/benchmark_failure_analysis.tex
\subsubsection{Failure analysis}
\label{sec:benchmark-failures}
We qualitatively review unsuccessful attempts to examine implementation errors in satisfying task specifications. We identify two main error patterns: (1) patches contain code or tests targeting a specification, yet their behavior still violates it. Incomplete handling of input boundaries or execution paths leaves the requirement satisfied only in some settings. (2) Some patches satisfy one requirement while violating another that constrains the same state. Although the model attempts to satisfy both requirements, it overlooks how a change made to satisfy one can violate the other in a different usage scenario. Concrete examples are provided in Appendix~\ref{app:failure-evidence}.

%% file: tables/rl_comparison.tex
\begin{table*}[t]
\centering
\begingroup
\small
\setlength{\tabcolsep}{5pt}
\renewcommand{\arraystretch}{1.14}
\begin{tabularx}{\linewidth}{
  >{\hsize=1.60\hsize\linewidth=\hsize\raggedright\arraybackslash}X
  >{\hsize=1.45\hsize\linewidth=\hsize\centering\arraybackslash}X
  >{\hsize=0.80\hsize\linewidth=\hsize\centering\arraybackslash}X
  >{\hsize=0.825\hsize\linewidth=\hsize\centering\arraybackslash}X
  >{\hsize=0.825\hsize\linewidth=\hsize\centering\arraybackslash}X
  >{\hsize=0.50\hsize\linewidth=\hsize\centering\arraybackslash}X}
\toprule
\textbf{Training Settings} & \textbf{\shortstack{\method-\\\textsc{Bench}}} & \textbf{\shortstack{DeepSWE-\\v1.1}} & \textbf{\shortstack{SWE-bench\\Verified}} & \textbf{\shortstack{SWE-bench\\Pro}} & \textbf{Avg} \\
\midrule
\multicolumn{6}{l}{\strut\textbf{MiMo-V2.5}} \\
w/ Vanilla & 47.00 & 12.39 & 69.27 & 50.11 & 44.69 \\
w/ SWE-PI & 44.00 & 6.78 & 68.00 & 49.43 & 42.05 \\
w/ MiMo Coding & \underline{50.00} & \underline{18.62} & \textbf{73.63} & \textbf{53.81} & \underline{49.01} \\
\rowcolor{tableours}
w/ \method{} & \textbf{52.00} & \textbf{25.30} & \underline{72.00} & \underline{51.30} & \textbf{50.15} \\
\midrule
\multicolumn{6}{l}{\strut\textbf{MiMo-V2.6-Flash-SFT}} \\
w/ Vanilla & 59.00 & 54.42 & 74.20 & 54.54 & 60.54 \\
w/ SWE-PI & 66.33 & \underline{56.64} & 76.60 & 59.05 & \underline{64.66} \\
w/ MiMo Coding & \underline{69.00} & 50.74 & \underline{77.20} & \textbf{59.44} & 64.09 \\
\rowcolor{tableours}
w/ \method{} & \textbf{70.67} & \textbf{62.98} & \textbf{77.40} & \underline{59.37} & \textbf{67.60} \\
\bottomrule
\end{tabularx}
\endgroup
\caption{Task success rates (\%, higher is better). Avg is the unweighted mean across the four benchmarks. Within each model, the best results are in \textbf{bold} and the second-best are \underline{underlined}.}
\label{tab:rl-matrix}
\end{table*}

%% file: tables/training_behavior.tex
\begin{table}[!t]
\centering\small
\setlength{\tabcolsep}{5pt}
\begin{tabularx}{\linewidth}{@{}Yrrrr@{}}
\toprule
Measure & \multicolumn{2}{c}{\textbf{MiMo-V2.5}} & \multicolumn{2}{c}{\textbf{MiMo-V2.6-Flash-SFT}}\\
\cmidrule(lr){2-3}\cmidrule(l){4-5}
& Vanilla & \method{} & Vanilla & \method{}\\
\midrule
Thinking tokens (k) & 29.7 & 141.6 & 22.8 & 69.5 \\
Modified files & 4.47 & 9.79 & 4.09 & 6.17 \\
Changed lines & 451 & 3,632 & 350 & 667 \\
Edit--test loops & 3.48 & 4.34 & 2.96 & 4.90 \\
Trajectories with $\geq 2$ loops (\%) & 63.0 & 72.7 & 69.0 & 78.3 \\
\bottomrule
\end{tabularx}
\caption{Reasoning, code changes, and edit--test loops on \bench{}. The edit--test loop definition is given in Appendix~\ref{app:development-behavior}.}
\label{tab:training-behavior}
\end{table}

%% file: sections/related_work.tex
\section{Related Work}
\label{sec:related-work}

\paragraph{Coding Data Synthesis.}
Prior work constructs coding tasks from source code, repository changes, and agent feedback.
Magicoder~\citep{magicoder} generates coding problems and solutions from open-source code snippets, while KodCode~\citep{kodcode} draws on programming exercises, algorithm implementations, and library documentation.
At the repository level, R2E-Gym~\citep{r2egym} derives tasks from commits through test generation and back-translation, SWE-smith~\citep{swesmith} modifies working code to create repair tasks, and SWE-Flow~\citep{sweflow} uses unit tests and runtime dependencies to construct incremental development tasks.
Other methods expand how task content is generated: SWE-Mirror~\citep{swemirror} transfers issue semantics across repositories, ADR~\citep{adr} recombines semantic elements extracted from seed problems, and TaskPilot~\citep{taskpilot} uses solver feedback to refine jointly generated tasks, reference patches, and tests.
\method{} makes the coverage and interaction of engineering requirements explicit objectives of task synthesis.
Guided by developer specification seeds and repository context, it constructs tasks that require joint reasoning about interdependent specifications.

\paragraph{Software Specifications.}
Software specifications make the behaviors and constraints of a development task explicit.
ArchCode~\citep{archcode} organizes functional and non-functional requirements from problem descriptions, while SpecRover~\citep{specrover} infers intended behavior from issue descriptions and repository context.
REAgent~\citep{reagent} produces structured requirements that include root-cause analysis and proposed edits, while CodeSpec~\citep{codespec} constructs cross-component functional chains to guide feature development.
Specifications also structure evaluation and learning: DevAI~\citep{devai} organizes development requirements as dependency graphs, SWE-AGI~\citep{sweagi} evaluates system construction from standards and RFCs, and SpecCoder~\citep{speccoder} trains models to produce specification analyses and code through supervised fine-tuning and reinforcement learning.
CodeSpec uses executable specifications to maintain consistency between a functional-chain design and its implementation.
\method{} instead uses interdependent specifications to construct the task: requirements are coupled so that the choices made to satisfy one constrain the valid choices for others.
The agent must explore the repository and develop an implementation that satisfies these requirements together.

%% file: sections/conclusion.tex
\section{Conclusion}
\label{sec:conclusion}

We introduced \method{}, a framework for synthesizing executable coding tasks with interdependent specifications in real repositories.
Starting from specification seeds drawn from developer interactions, a synthesis agent jointly refines task requirements and tests, while a review agent examines solutions from multiple models to identify verification errors.
The resulting training set contains 4,670 tasks across 2,865 repositories.
Reinforcement learning on these tasks substantially raises the base model's success rate on DeepSWE-v1.1, with gains over the base model on all four evaluated benchmarks.
We also introduced \bench{}, comprising 100 manually reviewed tasks and 1,044 specifications.
Evaluations reveal substantial performance differences across agents and remaining challenges in satisfying interacting requirements.

\section*{Limitations}

Our approach has three practical limitations.
First, using the synthesized tasks to improve model performance through RL requires substantial computational resources for training experiments.
Second, each task has an associated container image and runs in a Kubernetes pod.
Training and evaluation therefore require infrastructure to manage many containers concurrently.
Third, producing high-quality tasks relies on strong models and multiple rounds of filtering, which discard a large fraction of generated candidates.

%% file: sections/appendix.tex
\input{sections/appendix_method_data}
\input{sections/appendix_experiments}

\input{sections/appendix_analyses}
\input{sections/appendix_cases}

\input{sections/appendix_behavior_analysis}
\input{sections/appendix_failure_analysis}

%% file: sections/appendix_method_data.tex
\section{Data Construction}
\label{app:method-data}

\subsection{Agents}
\label{app:agents}

The synthesis and review agents use MiMo-V2.5-Pro.
Solver agents use MiMo-V2.5-Pro and MiMo-V2.5-Flash.
These agents use Claude Code~\citep{claude_code} as their harness.

\subsection{Manual review}
\label{app:author-review}

The authors manually reviewed the 100 tasks included in \bench{}, examining task descriptions, executable verifiers, code patches generated by models, and execution trajectories.
The review focused on whether tasks reflected realistic development scenarios in their repositories, whether requirements were clear and mutually consistent, and whether tests covered the explicitly stated requirements.
They also checked whether verifiers omitted required behavior or imposed constraints absent from the task descriptions, including implementation details that could lead to valid solutions being rejected.

The authors further examined successful and failed trajectories to assess the verification results.
For solutions that passed the tests, they checked whether the implementations actually satisfied the task requirements; for those that failed, they distinguished implementation defects from verifier restrictions and environment issues.
Solutions that were close to correct but failed verification helped identify hidden constraints and problems in handling boundary cases.
The review also considered missing dependencies, environment errors, and flaky tests to reduce their influence on the evaluation of model capabilities.

%% file: sections/appendix_experiments.tex
\section{Experimental Settings}
\label{app:experiments}

\input{sections/benchmark_evaluation_details}
\subsection{Reinforcement learning}
\label{app:training}

We train the model using GRPO~\citep{deepseekmath}. Table~\ref{tab:config} supplements the training setup in Section~\ref{sec:rl-utility}.

\begin{table}[htbp]
\centering
\small
\begin{tabular}{@{}lc@{}}
\toprule
\textbf{Setting} & \textbf{Value} \\
\midrule
Training batch size & 128 \\
Mini-batch size & 128 \\
Samples per task & 16 \\
Sampling temperature & 1.0 \\
Top-$p$ & 0.97 \\
Prompt token limit & 8,192 \\
Response token limit & 516,096 \\
\bottomrule
\end{tabular}
\caption{RL training hyperparameters.}
\label{tab:config}
\end{table}

\paragraph{Training dynamics.}
Training score averages within-task pass rates over all tested groups before batch selection, excluding invalid outcomes and assigning zero to groups with fewer than two valid outcomes.
Entropy is measured over masked training tokens; response length and main-agent turns are means per training context and sampled context, respectively.
The sampled tasks change across updates.

\subsection{Evaluation settings}
\label{app:rl-matrix}

The evaluation uses three samples per task, a 512K context budget, temperature 1.0, top-$p=0.95$.

%% file: sections/appendix_analyses.tex
\section{Specification Analysis}
\label{app:analyses}

\subsection{Categories and benchmark annotation}
\label{app:category-annotation}

Table~\ref{tab:spec-categories} summarizes the taxonomy across five primary categories, where functional semantics is further divided into three subcategories, yielding seven categories in total.

\input{tables/spec_categories}

We annotate the 1,044 specifications using a human-in-the-loop pipeline: MiMo-V2.5~\citep{mimo_v25} first proposes primary/secondary categories, auxiliary tags (compatibility, security, platform, observability), and rationales, which are subsequently audited by the authors. In Figure~\ref{fig:benchmark-spec-groups}, each specification is uniquely attributed to its primary category, while secondary categories track cross-cutting aspects.

Classification boundaries are grounded in the requirement's primary intent: interface contracts cover API signatures and schema constraints; functional semantics govern transformation logic; state obligations enforce invariants during failed transactions; and error handling captures explicit exceptions and retries. We further subdivide the 470 functional specifications into parsing, algorithmic computation, and domain protocols.

\subsection{Training query distributions}
\label{app:query-distributions}
\method{} queries have a median length of 2,131 characters, compared with 731--1,361 characters for the other training datasets.
Figure~\ref{fig:appendix-query-lengths} compares their distributions, and Table~\ref{tab:query-lengths} reports statistics over the complete training splits.

\par\medskip
\noindent
\begin{minipage}{\linewidth}
\captionsetup{hypcap=false}
\centering
\includegraphics[width=5.5in]{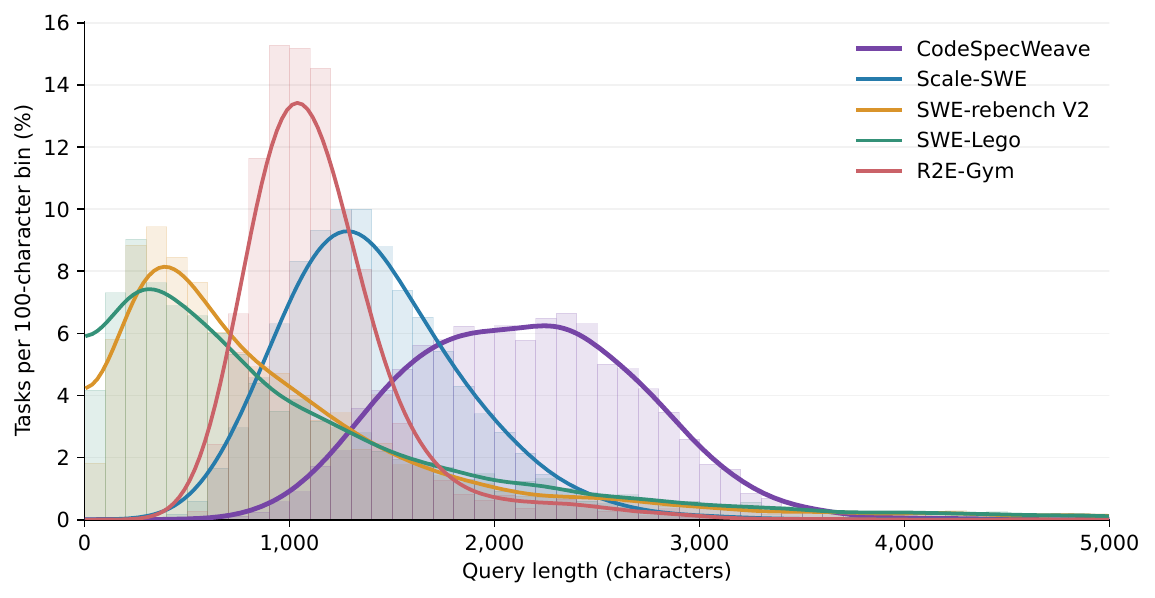}
\captionof{figure}{Query length distributions of \method{}, Scale-SWE, SWE-rebench V2, SWE-Lego, and R2E-Gym.}
\label{fig:appendix-query-lengths}

\vspace{10pt}
\input{tables/appendix_query_lengths}
\end{minipage}
\par\medskip

%% file: tables/spec_categories.tex
\begin{table}[!htbp]
\centering
\small
\begin{tabularx}{\linewidth}{@{}>{\raggedright\arraybackslash}p{0.36\linewidth}>{\raggedright\arraybackslash}X r@{}}
\toprule
Specification Type & Scope & Count \\
\midrule
\textbf{Functional Semantics} & End-to-end data transformation and algorithmic logic & 470 \\
\setlength{\leftskip}{0.6em}Parsing \& Representation & Syntax matching, encoding, and format serialization & 182 \\
\cmidrule[0.25pt](l{0.6em}){1-3}
\setlength{\leftskip}{0.6em}Computation \& Algorithms & Mathematical calculations and core algorithmic logic & 145 \\
\cmidrule[0.25pt](l{0.6em}){1-3}
\setlength{\leftskip}{0.6em}Domain \& Protocol Rules & Business constraints and protocol-specific semantics & 143 \\
\midrule
\textbf{Interface \& I/O Contracts} & Public APIs, argument types and defaults, input/output
schemas & 108 \\
\midrule[0.25pt]
\textbf{Control Flow \& Ordering} & Conditional branching, execution order, and concurrency & 98 \\
\midrule[0.25pt]
\textbf{State \& Side Effects} & State transitions, atomicity, and resource lifecycles & 181 \\
\midrule[0.25pt]
\textbf{Errors \& Recovery} & Input rejection, exception propagation, and fallback paths & 187 \\
\bottomrule
\end{tabularx}
\caption{Taxonomy and distribution of the 1,044 benchmark specifications. Functional specifications ($n=470$) are partitioned into three fine-grained subcategories.}
\label{tab:spec-categories}
\end{table}

%% file: tables/appendix_query_lengths.tex
\begingroup
\centering\small
\begin{tabular}{@{}lrrr@{}}\toprule
Dataset & Median & P10 & P90\\\midrule
\method{} & 2,131 & 1,414 & 2,856 \\
Scale-SWE & 1,361 & 900 & 2,006 \\
SWE-rebench V2 & 731 & 231 & 2,536 \\
SWE-Lego & 742 & 181 & 2,619 \\
R2E-Gym & 1,088 & 810 & 1,556 \\
\bottomrule\end{tabular}
\captionof{table}{Query lengths across the complete training splits, in characters.}
\label{tab:query-lengths}
\endgroup

%% file: sections/appendix_cases.tex
\section{Task Example}
\label{app:cases}
\label{sec:task-example}

This task illustrates how a new feature is constrained by compatibility and boundary requirements.
It asks to change how Skycoin's transaction builder allocates coin hours across payments while preserving its existing behavior for a single recipient.

\begin{casebox}{Task}
\small
\input{sections/skycoin_query.tex}
\end{casebox}

The requirements jointly constrain allocation, rounding, and output construction (Table~\ref{tab:task-requirements}).
The new proportional rule applies to multiple recipients, while the legacy cap must remain limited to a single recipient.
Applying both rules to every request would violate the new allocation policy.

\begin{table*}[!ht]
\centering\small

\begin{tabularx}{\linewidth}{@{}p{.22\linewidth}Y@{}}
\toprule
Requirement & Expected behavior\\
\midrule
Proportionality & Without change, allocate post-fee hours proportionally and resolve rounding ties by request order.\\
\midrule[0.25pt]
Change allocation & Split post-fee hours 50/50 between recipients and change, assigning the odd hour to change.\\
\midrule[0.25pt]
Minimum allocation & Give each recipient at least one hour whenever the recipient share permits it.\\
\midrule[0.25pt]
Output order & Keep recipients in request order and append non-zero-coin change last.\\
\midrule[0.25pt]
Compatibility & Preserve the single-recipient cap and its existing output values.\\
\bottomrule
\end{tabularx}
\caption{Requirements in the example task.}
\label{tab:task-requirements}
\end{table*}

%% file: sections/skycoin_query.tex
The CLI already supports building a raw transaction from a wallet through \texttt{createRawTransaction}/\texttt{send -m}, but its multi-recipient hour policy is not useful to callers: it divides recipient hours equally, can give a larger payment no more hours than a tiny payment, and can strand a recipient with zero hours when the available remainder is small. Make the raw-transaction builder allocate coin hours from the selected inputs according to the requested payments while preserving its existing transaction and wallet semantics.

For a non-empty multi-recipient request, after the normal required fee is reserved, allocate the recipient share in proportion to each recipient's coin amount using deterministic integer rounding: the recipient hours must sum exactly to the assignable recipient share, and any rounding remainder must be assigned deterministically without depending on map iteration or randomness; when fractional remainders tie, earlier entries in the request order receive the extra hours first. If there are enough recipient hours for every output to receive one, no recipient may receive zero. When a change output is present, split the post-fee hours between recipients and change using the existing 50/50 policy (the extra odd hour goes to change), then apply the proportional allocation only within the recipient share; all unassigned hours remain on the change output. When there is no change, do not burn or discard hours merely because the change output is absent. Keep output order stable: the \texttt{to} entries remain in request order and a non-zero-coin change output, if any, is last.

Do not regress the established single-recipient behavior: its existing coin-hour cap and output values must remain unchanged. Continue to reject requests that cannot produce a fee-valid transaction, and do not create zero-coin outputs.

%% file: sections/appendix_behavior_analysis.tex
\section{Development Behavior}
\label{app:new-analyses}
\label{app:development-behavior}

We identify editing and testing actions from tool names and shell commands in the main-agent trajectories.
An edit--test loop consists of one or more edits followed by a test invocation in a later turn.
Edits include source code, tests, and auxiliary files.
Within the first 40 turns, the proportion of trajectories with at least two edit--test loops changes from 44.0\% to 54.7\% for MiMo-V2.5 and from 47.7\% to 45.7\% for Flash-SFT.

\paragraph{Editing and retesting after failure.}
For test invocations whose output explicitly reports a failure, we measure how often a later edit is followed by another test invocation.
Both models more often edit and test again after \method{} training (Table~\ref{tab:failure-retest}).

\begin{table}[!htbp]
\centering\small
\begin{tabular}{@{}lrr@{}}
\toprule
Model  & Vanilla & \method{}\\
\midrule
MiMo-V2.5  & 77.1 & 86.0 \\
Flash-SFT  & 77.7 & 87.1 \\
\bottomrule
\end{tabular}
\caption{Task-averaged rates of editing and retesting after a failed test (\%). }
\label{tab:failure-retest}
\end{table}

%% file: sections/appendix_failure_analysis.tex
\section{Failure Analysis}
\label{app:failure-analysis}
\label{app:failure-evidence}

We use a frontier agent to inspect the prompt received by the solver, its final patch, and verification output, consulting test implementations and tool records where needed.
We first establish whether the prompt explicitly requires the failed behavior, then inspect the corresponding implementation and the execution evidence for its violation.

\begin{figure*}[p]
\centering
\includegraphics[width=\textwidth,height=.87\textheight,keepaspectratio]{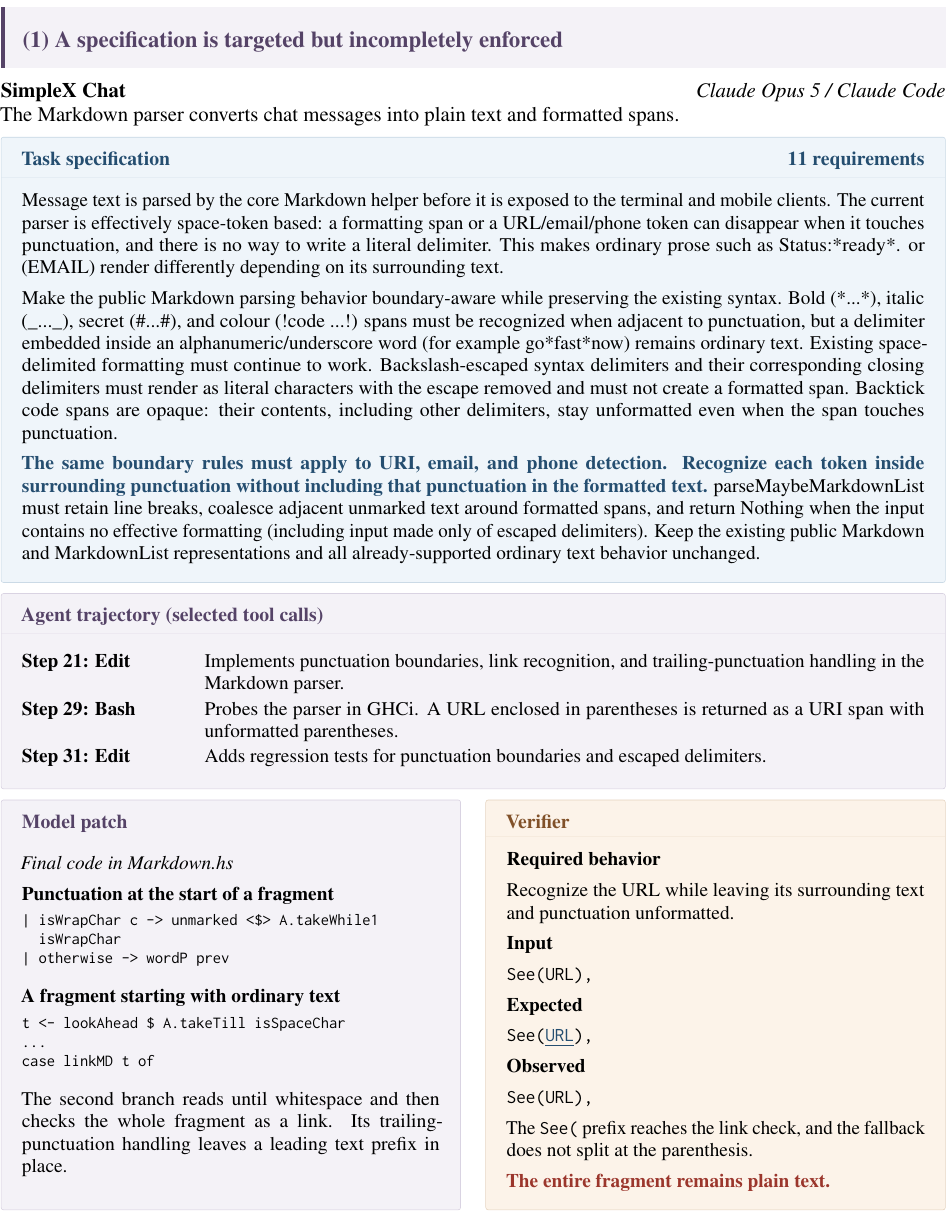}
\caption{\textbf{Failure pattern (1): a specification is targeted but incompletely enforced.} The patch includes link recognition and punctuation handling, and the agent's probe recognizes a parenthesized URL. Yet a URL following the text prefix remains plain text in the final evaluation. The targeted requirement is satisfied in one boundary setting but violated in another. URL and email strings are abbreviated; underlining denotes a URI span.}
\label{fig:failure-simplex}
\end{figure*}

\begin{figure*}[p]
\centering
\includegraphics[width=\textwidth,height=.87\textheight,keepaspectratio]{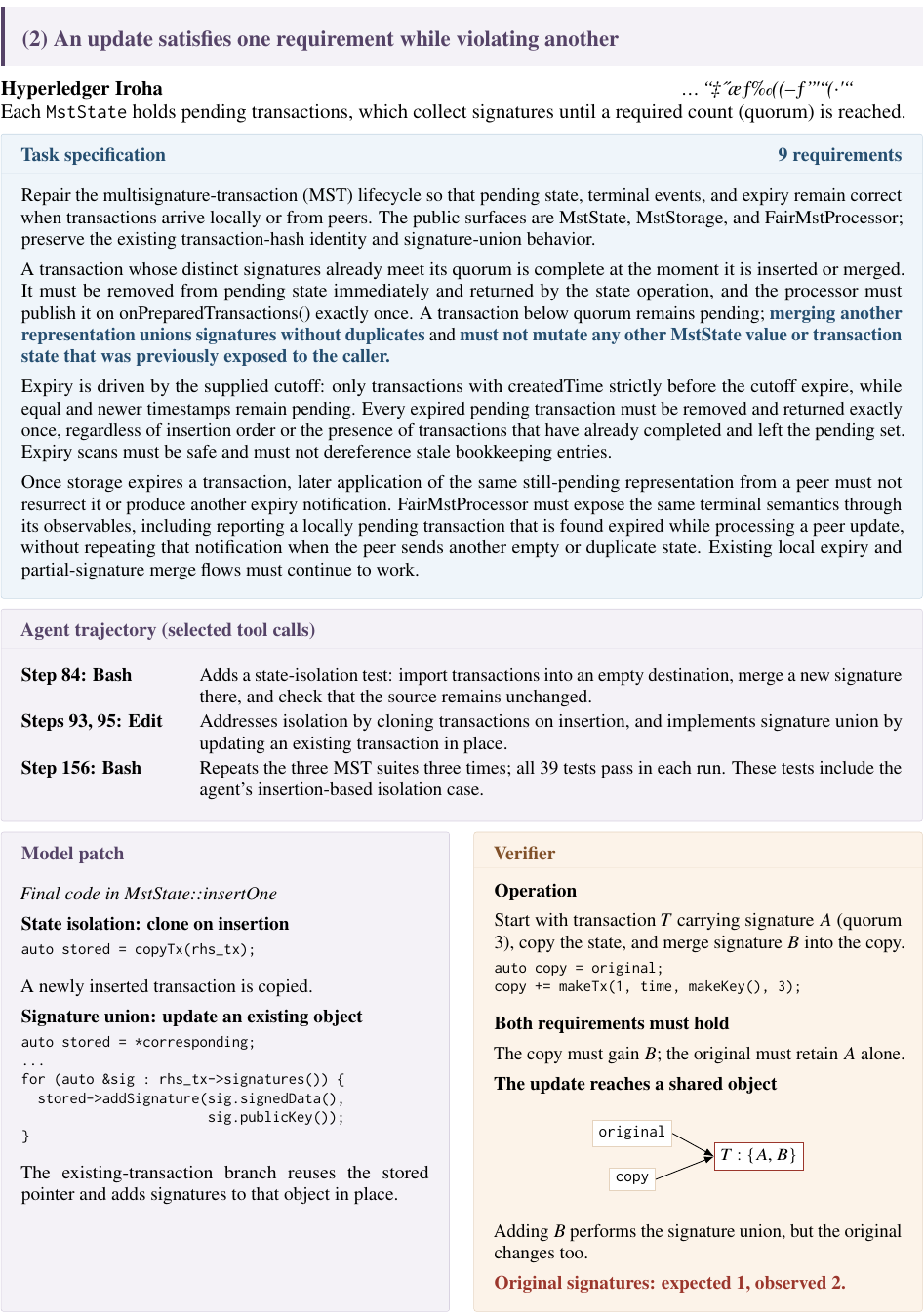}
\caption{\textbf{Failure pattern (2): an update satisfies one requirement while violating another.} The patch addresses both signature union and state isolation. After a whole-state copy, however, both states share a transaction. The update adds the required signature to the destination transaction while also changing the original, violating isolation. Thus the same update fails to satisfy both requirements together. The original-state count is measured by the verifier.}
\label{fig:failure-iroha}
\end{figure*}

Figures~\ref{fig:failure-simplex} and~\ref{fig:failure-iroha} illustrate error patterns (1) and (2) from Section~\ref{sec:benchmark-failures}, respectively.

\paragraph{Failure pattern (1): incomplete enforcement.}
The SimpleX patch in Figure~\ref{fig:failure-simplex} includes link recognition and punctuation handling, and the agent's probe recognizes a parenthesized URL.
However, a URL following \texttt{See(} remains plain text in the final evaluation.
The targeted boundary requirement therefore holds only in some input settings.

\paragraph{Failure pattern (2): interacting requirements.}
The Iroha patch in Figure~\ref{fig:failure-iroha} addresses both signature union and state isolation.
Transactions are cloned on insertion, but a whole-state copy retains shared transaction objects.
Adding the required signature through the copy also changes the original.
The same update therefore fulfills the merge requirement while violating isolation.

%% file: references.bib
@inproceedings{swebench,
  title={Swe-bench: Can language models resolve real-world github issues?},
  author={Jimenez, Carlos E and Yang, John and Wettig, Alexander and Yao, Shunyu and Pei, Kexin and Press, Ofir and Narasimhan, Karthik},
  booktitle={Proceedings of ICLR},
  volume={2024},
  pages={54107--54157},
  year={2024},
url = {https://proceedings.iclr.cc/paper_files/paper/2024/file/edac78c3e300629acfe6cbe9ca88fb84-Paper-Conference.pdf}
}

@inproceedings{swegym_icml,
   title = 	 {Training Software Engineering Agents and Verifiers with {SWE}-Gym},
  author =       {Pan, Jiayi and Wang, Xingyao and Neubig, Graham and Jaitly, Navdeep and Ji, Heng and Suhr, Alane and Zhang, Yizhe},
  booktitle = 	 {Proceedings of ICML},
  pages = 	 {47717--47737},
  year = 	 {2025},
  volume = 	 {267},
  url = 	 {https://proceedings.mlr.press/v267/pan25g.html},
}

@inproceedings{swesmith,
title={{SWE}-smith: Scaling Data for Software Engineering Agents},
author={John Yang and Kilian Lieret and Carlos E Jimenez and Alexander Wettig and Kabir Khandpur and Yanzhe Zhang and Binyuan Hui and Ofir Press and Ludwig Schmidt and Diyi Yang},
booktitle={Proceedings of NeurIPS Datasets and Benchmarks Track},
year={2026},
url={https://openreview.net/forum?id=63iVrXc8cC}
}

@inproceedings{sweflow,
  title = 	 {Synthesizing Software Engineering Data in a Test-Driven Manner},
  author =       {Zhang, Lei and Yang, Jiaxi and Yang, Min and Yang, Jian and Chen, Mouxiang and Zhang, Jiajun and Cui, Zeyu and Hui, Binyuan and Lin, Junyang},
  booktitle = 	 {Proceedings of ICML},
  pages = 	 {76518--76540},
  year = 	 {2025},
  volume = 	 {267},
  publisher =    {PMLR},
  url = 	 {https://proceedings.mlr.press/v267/zhang25cn.html}
}

@article{r2egym,
  title={R2e-gym: Procedural environments and hybrid verifiers for scaling open-weights swe agents},
  author={Jain, Naman and Singh, Jaskirat and Shetty, Manish and Zheng, Liang and Sen, Koushik and Stoica, Ion},
  journal={arXiv preprint arXiv:2504.07164},
 url = {https://arxiv.org/abs/2504.07164},
  year={2025}
}

@article{datacurve_v11,
  title={DeepSWE: Measuring Frontier Coding Agents on Original, Long-Horizon Engineering Tasks},
  author={Huang, Wenqi and Lee, Charley and Tng, Leonard and Ge, Serena},
  journal={arXiv preprint arXiv:2607.07946},
   url = {https://arxiv.org/abs/2607.07946},
  year={2026}
}

@inproceedings{swemutation,
    title = "{SWE}-Mutation: Can {LLM}s Generate Reliable Test Suites in Software Engineering?",
    author = "Sun, Yuxuan  and
      Zhao, Yuze  and
      Wang, Yufeng  and
      Du, Yao  and
      Ma, Zhiyuan  and
      Wang, Jinbo  and
      Zhang, Mengdi  and
      Zhang, Kai  and
      Huang, Zhenya",
    booktitle = "Findings of ACL",
    year = "2026",
    url = "https://aclanthology.org/2026.findings-acl.1976/",
    pages = "39651--39674",
}

@article{coharden,
  title={Beyond Fail-to-Pass: Iterative Hardening of Co-Generated Bug Reproduction Tests and Fixes},
  author={Tan, Yuhao and Yang, Zhibang and Yang, Fangkai and Yao, Yuan and Kang, Yu and Wang, Lu and Zhao, Pu and Zhang, Xin and Ma, Xiaoxing and Lin, Qingwei and others},
  journal={arXiv preprint arXiv:2607.19843},
  year={2026},
  url = {https://arxiv.org/abs/2607.19843}
}

@inproceedings{
swebench_pro,
title={{SWE}-Bench Pro: Can {AI} Agents Solve Long-Horizon Software Engineering Tasks?},
author={Xiang Deng and Jeff Da and Edwin Pan and Yannis Yiming He and Charles Ide and Kanak Garg and Niklas Lauffer and Andrew Park and Chetan Rane and Karmini Sampath and Maya Krishnan and Srivatsa R Kundurthy and Sean M. Hendryx and Zifan Wang and Chen Bo Calvin Zhang and Noah Jacobson and Bing Liu and Brad Kenstler},
booktitle={Proceedings of ICML},
year={2026},
url={https://openreview.net/forum?id=uEVTdoAbnK}
}

@inproceedings{archcode,
    title = "{A}rch{C}ode: Incorporating Software Requirements in Code Generation with Large Language Models",
    author = "Han, Hojae  and
      Kim, Jaejin  and
      Yoo, Jaeseok  and
      Lee, Youngwon  and
      Hwang, Seung-won",
    booktitle = "Proceedings of ACL",
    year = "2024",
    url = "https://aclanthology.org/2024.acl-long.730/",
    pages = "13520--13552",
}

@article{adr,
  title={Combinatorial Synthesis: Scaling Code RLVR via Atomic Decomposition and Recombination},
  author={Zheng, Jiasheng and Cao, Boxi and Yu, Boxi and Zhang, Yuzhong and Cao, Jialun and Lu, Yaojie and Lin, Hongyu and Han, Xianpei and Sun, Le},
  journal={arXiv preprint arXiv:2605.31058},
  url = {https://arxiv.org/abs/2605.31058},
  year={2026}
}

@article{codespec,
  title={CodeSpec: Dual Executable Specifications for Agentic Long-Horizon Feature Development},
  author={Wang, Peiding and Zhang, Li and Liu, Fang and Li, Taichuan and Zhu, Yinghao},
  journal={arXiv preprint arXiv:2607.26777},
  year={2026},
  url = {https://arxiv.org/abs/2607.26777}
}

@inproceedings{kodcode,
    title = "{K}od{C}ode: A Diverse, Challenging, and Verifiable Synthetic Dataset for Coding",
    author = "Xu, Zhangchen  and
      Liu, Yang  and
      Yin, Yueqin  and
      Zhou, Mingyuan  and
      Poovendran, Radha",
    booktitle = "Findings of ACL",
    year = "2025",
    url = "https://aclanthology.org/2025.findings-acl.365/",
    pages = "6980--7008",
}

@inproceedings{devai,
  title = 	 {Agent-as-a-Judge: Evaluate Agents with Agents},
  author =       {Zhuge, Mingchen and Zhao, Changsheng and Ashley, Dylan R. and Wang, Wenyi and Khizbullin, Dmitrii and Xiong, Yunyang and Liu, Zechun and Chang, Ernie and Krishnamoorthi, Raghuraman and Tian, Yuandong and Shi, Yangyang and Chandra, Vikas and Schmidhuber, J\"{u}rgen},
  booktitle = 	 {Proceedings of ICML},
  pages = 	 {80569--80611},
  year = 	 {2025},
  volume = 	 {267},
  month = 	 {13--19 Jul},
  url = 	 {https://proceedings.mlr.press/v267/zhuge25a.html},
}

@article{swemirror,
  title={Swe-mirror: Scaling issue-resolving datasets by mirroring issues across repositories},
  author={Wang, Junhao and Zan, Daoguang and Xin, Shulin and Liu, Siyao and Wu, Yurong and Shen, Kai},
  journal={arXiv preprint arXiv:2509.08724},
  year={2025},
  url = {https://arxiv.org/abs/2509.08724}
}

@article{reagent,
  title={Reagent: Requirement-driven llm agents for software issue resolution},
  author={Kuang, Shiqi and Tian, Zhao and Lin, Kaiwei and Tao, Chaofan and Wang, Shaowei and Bai, Haoli and Shang, Lifeng and Chen, Junjie},
  journal={arXiv preprint arXiv:2604.06861},
  year={2026},
  url = {https://arxiv.org/abs/2604.06861}
}

@article{taskpilot,
  title={FrogNano: Training a 4B Coding Agent via Online Task Synthesis},
  author={Kim, Minseon and Shi, Zhengyan and Penaloza, Emiliano and Cui, Christopher and Castanyer, Roger Creus and Hashemzadeh, Maryam and White, Isadora and Light, Jonathan and Kim, Jeonghye and Pereira, Matheus and others},
  journal={arXiv preprint arXiv:2609.07925},
  year={2026},
  url = {https://arxiv.org/abs/2609.07925}
}

@article{speccoder,
  title={SpecCoder: Specification-Aware Code Generation with Curriculum Dual-Task Reinforcement Learning},
  author={Li, Yixuan and Huang, Mingxuan and Wang, Jiajing and Yang, Weidong and Liu, Xinyi and Fei, Ben and Ma, Lipeng},
  journal={arXiv preprint arXiv:2609.06041},
  year={2026},
  url = {https://arxiv.org/abs/2609.06041}
}

@inproceedings{specrover,
  title={Specrover: Code intent extraction via llms},
  author={Ruan, Haifeng and Zhang, Yuntong and Roychoudhury, Abhik},
  booktitle={Proceedings of ICSE},
  pages={963--974},
  year={2025}
}

@InProceedings{magicoder,
  title = 	 {Magicoder: Empowering Code Generation with {OSS}-Instruct},
  author =       {Wei, Yuxiang and Wang, Zhe and Liu, Jiawei and Ding, Yifeng and Zhang, Lingming},
  booktitle = 	 {Proceedings of ICML},
  pages = 	 {52632--52657},
  year = 	 {2024},
  editor = 	 {Salakhutdinov, Ruslan and Kolter, Zico and Heller, Katherine and Weller, Adrian and Oliver, Nuria and Scarlett, Jonathan and Berkenkamp, Felix},
  volume = 	 {235},
  month = 	 {21--27 Jul},
  url = 	 {https://proceedings.mlr.press/v235/wei24h.html},
}

@article{sweagi,
  title={SWE-AGI: Benchmarking specification-driven software construction with MoonBit in the era of autonomous agents},
  author={Zhang, Zhirui and Zhang, Hongbo and Fei, Haoxiang and Bao, Zhiyuan and Chen, Yubin and Lei, Zhengyu and Liu, Ziyue and Sun, Yixuan and Xiao, Mingkun and Ye, Zihang and others},
  journal={arXiv preprint arXiv:2602.09447},
  url = {https://arxiv.org/abs/2602.09447},
  year={2026}
}

@inproceedings{sweagent,
 author = {Yang, John and Jimenez, Carlos and Wettig, Alexander and Lieret, Kilian and Yao, Shunyu and Narasimhan, Karthik and Press, Ofir},
 booktitle = {Proceedings of NeurIPS},
 pages = {50528--50652},
 title = {SWE-agent: Agent-Computer Interfaces Enable Automated Software Engineering},
 url = {https://proceedings.neurips.cc/paper_files/paper/2024/file/5a7c947568c1b1328ccc5230172e1e7c-Paper-Conference.pdf},
 volume = {37},
 year = {2024}
}

@article{deepseekmath,
  title={Deepseekmath: Pushing the limits of mathematical reasoning in open language models},
  author={Shao, Zhihong and Wang, Peiyi and Zhu, Qihao and Xu, Runxin and Song, Junxiao and Bi, Xiao and Zhang, Haowei and Zhang, Mingchuan and Li, YK and Wu, Yang and others},
  journal={arXiv preprint arXiv:2402.03300},
  year={2024},
  url = {https://arxiv.org/abs/2402.03300}
}

@misc{swebench_verified,
  title = {{Introducing SWE-bench Verified}},
  author = {Chowdhury, Neil and Aung, James and Shern, Chan Jun and Jaffe, Oliver and Sherburn, Dane and Starace, Giulio and Mays, Evan and Dias, Rachel and Aljubeh, Marwan and Glaese, Mia and Jimenez, Carlos E. and Yang, John and Ho, Leyton and Patwardhan, Tejal and Liu, Kevin and Madry, Aleksander},
  year = {2024},
  month = aug,
  howpublished = {OpenAI blog},
  note = {Published August 13, 2024},
  url = {https://openai.com/index/introducing-swe-bench-verified/}
}

@misc{gpt54,
  title = {{GPT-5.4 Thinking System Card}},
  author = {{OpenAI}},
  year = {2026},
  month = mar,
  howpublished = {System card},
  url = {https://deploymentsafety.openai.com/gpt-5-4-thinking}
}

@misc{gpt55,
  title = {{GPT-5.5 System Card}},
  author = {{OpenAI}},
  year = {2026},
  month = apr,
  howpublished = {System card},
  url = {https://deploymentsafety.openai.com/gpt-5-5}
}

@misc{gpt56sol,
  title = {{GPT-5.6 System Card}},
  author = {{OpenAI}},
  year = {2026},
  month = jul,
  howpublished = {System card},
  url = {https://deploymentsafety.openai.com/gpt-5-6}
}

@misc{gpt6astra,
  title = {{GPT-6 Astra System Card}},
  author = {{OpenAI}},
  year = {2026},
  month = sep,
  howpublished = {System card},
  url = {https://deploymentsafety.openai.com/gpt-6-astra}
}

@misc{claude_opus5,
  title = {{System Card: Claude Opus 5}},
  author = {{Anthropic}},
  year = {2026},
  month = jul,
  howpublished = {System card},
  url = {https://www.anthropic.com/claude-opus-5-system-card}
}

@misc{kimi_k3,
  title = {{Kimi K3}},
  author = {{Moonshot AI}},
  year = {2026},
  howpublished = {Hugging Face model card},
  url = {https://huggingface.co/moonshotai/Kimi-K3}
}

@misc{qwen38max,
  title = {{Qwen3.8-Max: A New Bar for Coding and Cowork}},
  author = {{Qwen Team}},
  year = {2026},
  month = aug,
  howpublished = {Qwen blog},
  url = {https://qwen.ai/blog?id=qwen3.8}
}

@misc{mimo_v25,
  title = {{MiMo-V2.5}},
  author = {{Xiaomi MiMo Team}},
  year = {2026},
  howpublished = {Hugging Face model card},
  url = {https://huggingface.co/XiaomiMiMo/MiMo-V2.5}
}

@misc{mimo_v25_pro,
  title = {{MiMo-V2.5-Pro}},
  author = {{Xiaomi MiMo Team}},
  year = {2026},
  howpublished = {Hugging Face model card},
  url = {https://huggingface.co/XiaomiMiMo/MiMo-V2.5-Pro}
}

@misc{mimo_v26,
  title = {{MiMo-V2.6: Scaling Reinforcement Learning Towards Self-Improvement}},
  author = {{LLM-Core Xiaomi}},
  year = {2026},
  howpublished = {Technical report},
  url = {https://huggingface.co/XiaomiMiMo/MiMo-V2.6-Flash-RL/resolve/main/MiMo_V2_6_technical_report.pdf}
}

@misc{codex_cli,
  title = {{Codex CLI}},
  author = {{OpenAI}},
  url = {https://learn.chatgpt.com/docs/codex/cli}
}

@misc{claude_code,
  title = {{Claude Code Overview}},
  author = {{Anthropic}},
  url = {https://code.claude.com/docs/en/overview}
}

@article{scaleswe,
  title={Immersion in the github universe: Scaling coding agents to mastery},
  author={Zhao, Jiale and Chen, Guoxin and Meng, Fanzhe and Li, Minghao and Chen, Jie and Xu, Hui and Sun, Yongshuai and Zhao, Wayne Xin and Song, Ruihua and Zhang, Yuan and others},
  journal={arXiv preprint arXiv:2602.09892},
url = {https://arxiv.org/abs/2602.09892},
  year={2026}
}

@article{swelego,
  title={Swe-lego: Pushing the limits of supervised fine-tuning for software issue resolving},
  author={Tao, Chaofan and Chen, Jierun and Jiang, Yuxin and Kou, Kaiqi and Wang, Shaowei and Wang, Ruoyu and Li, Xiaohui and Yang, Sidi and Du, Yiming and Dai, Jianbo and others},
  journal={arXiv preprint arXiv:2601.01426},
    url = {https://arxiv.org/abs/2601.01426},
  year={2026}
}

@inproceedings{
swerebench,
title={{SWE}-rebench V2: Language-Agnostic {SWE} Task Collection at Scale},
author={Ibragim Badertdinov and Maksim Nekrashevich and Anton Shevtsov and Alexander Golubev},
booktitle={Proceedings of ICML},
year={2026},
url={https://openreview.net/forum?id=UCAda9kS57}
}

@misc{primeintellect2026scalingagenticrl,
  author = {Auras, Daniel and {Prime Intellect Team}},
  title = {{Scaling Agentic RL: 365,000+ Environments for SWE, Terminal, and Search}},
  howpublished = {Prime Intellect blog},
  year = {2026},
  month = jul,
  url = {https://www.primeintellect.ai/blog/scaling-agentic-rl}
}

@misc{multiswebench,
  title = {{Multi-SWE-bench: A Multilingual Benchmark for Issue Resolving}},
  author = {Daoguang Zan and Zhirong Huang and Wei Liu and Hanwu Chen and Linhao Zhang and Shulin Xin and Lu Chen and Qi Liu and Xiaojian Zhong and Aoyan Li and Siyao Liu and Yongsheng Xiao and Liangqiang Chen and Yuyu Zhang and Jing Su and Tianyu Liu and Rui Long and Kai Shen and Liang Xiang},
  year = {2025},
  eprint = {2504.02605},
  archivePrefix = {arXiv},
  primaryClass = {cs.SE},
  url = {https://arxiv.org/abs/2504.02605}
}

@article{tan2026swe,
  title={{SWE-Touch}: Benchmarking Coding Agents When Users Touch the Code},
  author={Tan, Yuqiao and Meng, Jinxiang and Lei, Fangyu and Wang, Minzheng and He, Shizhu and Zhao, Jun and Liu, Kang},
  journal={arXiv preprint arXiv:2608.02499},
  url={https://arxiv.org/abs/2608.02499},
  year={2026}
}
